\documentclass[jmp]{revtex4-1}

\usepackage{amsfonts}
\usepackage{makeidx}
\usepackage{amsmath,graphicx}
\usepackage{latexsym,mathrsfs}
\usepackage{amsmath,graphicx}
\usepackage{latexsym,mathrsfs}
\usepackage{amssymb}
\usepackage{eucal}

\makeindex

\def\Res{\mathop{\mbox{Res}\,}\limits}

\def\const{\mathrm{const\,}}

\allowdisplaybreaks
\def\openone{\leavevmode\hbox{\small1\kern-3.3pt\normalsize1}}

\def\a{{\boldsymbol a}}

\def\openone{\leavevmode\hbox{\small1\kern-3.3pt\normalsize1}}
\def\openone{\leavevmode\hbox{\small1\kern-3.3pt\normalsize1}}

\def\a{{\boldsymbol a}}

\def\0{{\boldsymbol 0}}

\def\Res{\mathop{\mbox{Res}\,}\limits}

\begin{document}

\title{\textbf{Exact solutions for a class of integrable
H\'{e}non-Heiles-type systems}%
}
\author{N. A. Kostov}
\email{nakostov@inrne.bas.bg}
\author{V. S. Gerdjikov}
\email{gerjikov@inrne.bas.bg}
\affiliation{Institute for Nuclear Research and Nuclear Energy,\\
Bulgarian Academy of Sciences, \\
72 Tsarigradsko chaussee 1784 Sofia, Bulgaria}
\author{V. Mioc}
\email{vasile_mioc@yahoo.com}
\affiliation{Astronomical Institute of the Romanian Academy, \\
Str. Cutitul de Argint 5, \\
RO-040557 Bucharest, Romania}

\date{\today}

\begin{abstract}
We study the exact solutions of a class of integrable
H\'{e}non-Heiles-type systems (according to the analysis of
Bountis et al. (1982)). These solutions are expressed in terms of
two-dimensional Kleinian functions. Special periodic solutions are
expressed in terms of the well-known Weierstrass function. We
extend some of our results to a generalized H\'{e}non-Heiles-type system
with  $(n+1)$ degrees of freedom.
\end{abstract}



\maketitle

\section{Introduction}
The famous H\'{e}non-Heiles potential (for details, see Section 2)
was introduced to model the motion of a star within a galaxy. But
it has much larger connotations.

Mathematically speaking, the model was constructed by adding two
terms of third degree to the potential of a planar oscillator. Such a
mathematical model also
appears by expanding the potential corresponding to an integrable
system (resulting via some canonical transformations applied to
the motion of three particles on a circle under exponentially
decreasing forces) to third degree terms (Boccaletti and Pucacco
1996 \cite{bp96}; Anisiu and Pal 1999 \cite{ap99}).

This problem originates in a question largely discussed in the sixties and
approached much earlier: the existence of a third isolating integral of the
motion (besides the integrals of energy and angular momentum). For terminology,
significance, and endeavours , see (H\'{e}non and Heiles 1964 \cite{HeHe} and
the references therein).

Starting with 1957, G. Contopoulos searched for and found some cases of
two potentials when such an integral ( $I_3$, in the classical notation) exists.
Quoting him \cite{Con,Conto}, numerical results provide ample evidence of
the existence of $I_3$ in quite general potential fields. As concrete exemplifications
from astronomy, but not only: slightly elliptical stellar clusters or galaxies,
Schmidt's model of the Galaxy, noncentral gravitational field of the Earth;
also, systems where the unperturbed frequency ratio is rational, or potential fields
deprived of a symmetry plane.

Besides these situations, there exist problems of statistical mechanics \cite{Ford,WF,Coh},
celestial mechanics \cite{Win} and quantum mechanics \cite{NKM} that join a Henon-Heiles-type model.

In 1964 M. H\'{e}non and C. Heiles \cite{HeHe} simplified the initial problem and approached
it numerically via Poincar\'{e} sections. Their paper and their numerical experiments
had a great echo, benefitting today of more than 600 citations in the mathematical,
physical and astronomical literature. The model bears their names, and the generalizations
are called H\'{e}non-Heiles-type models.

As a matter of fact, in many H\'{e}non-Heiles-type models (and earlier models) the
integral of angular momentum does not hold. Such models were also
tackled via the tools of the theory of dynamical systems. Only one
example at hand: the study of the collision and escape dynamics in
the associated two-body problem \cite{MB*03, MPC*09}.

Coming back to the H\'{e}non-Heiles-type model detailed in Section \ref{2}, in 1982,
T. Bountis et al. \cite{BSV} used Painlev\'{e} analysis to
identify all integrable cases. They identified three such
situations, ruled by the values of the free parameters of the
model; out of them case (ii) is the most general (see Section \ref{2}).

The present paper intends to point out exact solutions for the
class of H\'{e}non- Heiles-type systems described by the case (ii)
above. Section \ref{2} establishes the basic equations for the problem
to be investigated. We start from cylindrical coordinates (in the
motion reduced to a meridian plane), but then we work with a
generalized H\'{e}non-Heiles-type system in configuration-momentum
coordinates. In Section \ref{3} we derive the Lax representation for the
corresponding system.

Section \ref{4} provides exact quasi-periodic solutions for the system
under consideration. These orbits are
expressed in terms of Kleinian hyperelliptic functions.

In Section \ref{5} we search for elliptic periodic solutions of our system. To this
end in view, we resort to the method proposed in \cite{bbeim94,ek94,ee94b}, which
allows the construction of periodic solutions in a straightforward way, applying
the spectral theory for the Schr\"{o}dinger equation to elliptic potentials. In this
way we get the corresponding solution of the system under consideration.

Section \ref{6} tackles a generalized
H\'{e}non-Heiles-type system with $(n+1)$ degrees of freedom. We
apply the methods presented in the previous sections to such a
system, and point out the exact solutions in this case.

The final
Section \ref{7} summarizes the main results obtained in the paper and
formulates some conclusions.

\section{Basic equations}
\label{2}
We consider here a model for the dynamics of a point mass (a star) within
the gravitational field ruled by the potential $V_{g}$ of an axially
symmetric galaxy. Let the mass of the star be $m$, and let the cylindrical
coordinates we shall use be $(r,\phi ,z)$. The axis $Oz$ is the axis of
symmetry, $z$ is the distance of the star from the reference plane, $r:=%
\sqrt{x^{2}+y^{2}}$ is the distance between the star and the axis $Oz$ and $%
\phi :=\arctan \frac{y}{x}$ is the polar angle. Here the $Oxy$-plane is a
reference plane (in rectangular coordinates) that defines the coordinate $r$
and the angle $\phi $.

The Hamiltonian of the model can be written as
\begin{equation*}
H(p_{r},p_{z},p_{\phi },r,z,\phi )=\frac{1}{2m}\left(
p_{r}^{2}+p_{z}^{2}\right) +\frac{1}{2mr}p_{\phi }^{2}+V_{g}(r,z),
\end{equation*}%
where $p_{r}=m\dot{r}$ and $p_{z}=m\dot{z}$ are the linear momenta in the $r$
and $z$ directions respectively, $p_{\phi }=mr\dot{\phi}$ is the angular
momentum around the symmetry axis, whereas $\dot{}=\frac{d}{dt}$ is the
derivative with respect to the time $t$.

We have two integrals of motion, the total energy, $E$, and the angular
momentum, $l$, respectively:
\begin{eqnarray}
E &=&V_{g}(r,z)+\tfrac{m}{2}\big(\dot{r}^{2}+r^{2}\dot{\phi}^{2}+\dot{z}^{2}%
\big), \\[4pt]
l &=&mr\dot{\phi}.
\end{eqnarray}%
With the help of the second integral $l$, we reduce the dynamics of the star
to the meridian plane $(r,z)$:
\begin{eqnarray}
&&{\displaystyle\ddot{r}=-\frac{\partial V(r,z)}{\partial r}}\,,\qquad {%
\displaystyle\ddot{z}=-\frac{\partial V(r,z)}{\partial z}}\,,  \label{hh} \\%
[4pt]
&&V(r,z)\ :=\ V_{g}(r,z)+\tfrac{m}{2}r^{2}\dot{\phi}^{2}\ =\ V_{g}(r,z)+%
\frac{l^{2}}{2mr^{2}}.  \notag
\end{eqnarray}

This problem originates in a question discussed in the sixties: the
existence of a third isolating integral of the motion ($I_{3}$), see the
introductory section. Starting with 1957, G.\thinspace Contopoulos \cite{Con}
found some cases of potentials $V(r,z)$ when such integral $I_{3}$ exists.

In 1964 M.\thinspace H\'{e}non and C.\thinspace Heiles \cite{HeHe}
simplified the problem, canceling all terms of order $\geq 4$ in the
potential $V(r,z)$. The simplest such potential can be read as
\begin{equation*}
V(r,z)=Cr^{3}+\frac{1}{2}r z^{2} +\frac{1}{2}\left( Ar^{2}+Bz^{2}\right) ,
\end{equation*}%
where $A,B$ and $C$ are free parameters.

Using Painlev\'{e} analysis, in 1982 T.\thinspace Bountis, H.\thinspace
Segur and F.\thinspace Vivaldi \cite{BSV} extracted all integrable cases of (%
\ref{hh}), see also \cite{ctw82,dgr83,rgb89}:
\begin{eqnarray}
&\mbox{(i)}&\qquad A=B,\ C=\frac{1}{3} \\[5pt]
&\mbox{(ii)}&\qquad C=1,{\mbox{\ A\ and\ B\ are\ arbitrary}} \\[5pt]
&\mbox{(iii)}&\qquad 16A=B,\ C=\frac{16}{3}\,.
\end{eqnarray}%
In the case (i), the equations of motion decoupled in $(z+r)$, $(z-r)$
coordinates and the general solution can be expressed via elliptic
functions, see for example \cite{vmc02} and references therein.
The question about the effective exact solution in the case (iii)
is still open \cite{vmc02}. For the case (ii) effective solutions are obtained in \cite%
{Gav,rgc93,zm01,Mak}. In this paper we shall focus our attention to the H\'{e}%
non-Heiles-type systems related to second case (ii).

Next we will use ${\tilde q}_{1}$ and ${\tilde q}_{2}$ instead of $r$ and $z$. We consider a
generalized H\'{e}non-Heiles-type system with two degrees of freedom \cite%
{f91,aw92,t94,t95,mt97} :
\begin{eqnarray}
&&\ddot{q}_{1}+3q_{1}^{2}+\frac{1}{2}q_{2}^{2}+a_{0}q_{1}-\frac{a_{1}}{4}=0,
\label{systemH12} \\
&&\ddot{q}_{2}+q_{1}q_{2}-\frac{a_{4}}{4q_{2}^{3}}+\frac{a_{0}}{4}q_{2}=0.
\label{systemH22}
\end{eqnarray}%
Its Hamiltonian is
\begin{equation}
H_{0}=\frac{1}{2}(p_{1}^{2}+p_{2}^{2})+q_{1}^{3}+\frac{1}{2}q_{1}q_{2}^{2}+%
\frac{a_{4}}{8q_{2}^{2}}+\frac{a_{0}}{2}\left( q_{1}^{2}+\frac{1}{4}%
q_{2}^{2}\right) -\frac{a_{1}}{4}q_{1},  \label{HH7}
\end{equation}%
where $q_{1},q_{2},p_{1},p_{2}$ are the canonical coordinates and momenta
and $a_{0},a_{1},a_{4}$ are free constant parameters. Moreover $H_{0}$ is
related to the Hamiltonian
\begin{equation}
H_{H}=\frac{1}{2}(p_{1}^{2}+p_{2}^{2})+\bar{q}_{1}^{3}+\frac{1}{2}\bar{q}_{1}%
\bar{q}_{2}^{2}+\frac{\bar{a_{4}}}{{8\bar{q}_{2}^{2}}}+\frac{1}{2}\left( A%
\bar{q}_{1}^{2}+B\bar{q}_{2}^{2}\right) ,
\end{equation}%
through the map
\begin{eqnarray}
&&q_{1}=\bar{q}_{1}+\frac{A}{2}-2B,\quad q_{2}=\bar{q}_{2},  \notag \\
&&a_{0}=-2A+12B,\quad a_{1}=-A^{2}+16AB-48B^{2}.
\end{eqnarray}%
The function $H_{H}$ is the Hamiltonian of a classical integrable H\'{e}%
non-Heiles system with the additional term $a_{4}/8\bar{q}_{2}^{2}$. The
corresponding equations are equivalent to the ordinary differential equation
for travelling wave solutions of the fifth-order flow in the Korteweg - de
Vries (KdV) hierarchy \cite{f91}.

\section{Lax representation}
\label{3}
Next we will derive $(2\times 2)$ matrix Lax representation for the
generalized H\'{e}non-Heiles system (\ref{HH7}). The Lax representation has
the form \cite{m80,m84}
\begin{equation}
\dot{L}=[M(t,\lambda ),L(t,\lambda )],\quad L=\left(
\begin{array}{cc}
V & U \\
W & -V%
\end{array}%
\right) ,\quad M=\left(
\begin{array}{cc}
0 & 1 \\
Q & 0%
\end{array}%
\right)  \label{lax3}
\end{equation}%
where $U,W,Q$ are \cite{eekl93,eekt94,hkr99} see also \cite{k02}:
\begin{eqnarray}
U(t,\lambda )&=&F(t,\lambda )=\lambda ^{2}+\frac{1}{2}q_{1}\lambda -\frac{1}{%
16}q_{2}^{2},  \notag \\
V(t,\lambda )&=&-\frac{1}{2}\dot{F}(t,\lambda )=-\frac{1}{4}p_{1}\lambda +%
\frac{1}{16}q_{2}p_{2},  \notag \\
W(t,\lambda )&=&-\frac{1}{2}\ddot{F}+QF=\lambda ^{3}-(\frac{1}{2}q_{1}+\frac{%
1}{4}a_{0})\lambda ^{2}  \notag \\
&+&(\frac{1}{4}q_{1}^{2}+\frac{1}{16}q_{2}^{2}-\frac{1}{16}a_{1}+\frac{1}{8}%
a_{0}q_{1})\lambda +\frac{1}{16}p_{2}^{2}+\frac{1}{64}\frac{a_{4}}{q_{2}^{2}}%
,  \notag \\
Q(t,\lambda )&=&\lambda -q_{1}-\frac{1}{4}a_{0}.  \notag
\end{eqnarray}%
The corresponding algebraic curve is $\det({\mbox
L(t,\lambda)-\frac{\nu}{2} I })=0$
\begin{eqnarray}\label{curvecan1}
\nu ^{2}=4\lambda ^{5}-a_{0}\lambda ^{4}-\frac{1}{4}a_{1}\lambda ^{3}+\frac{1%
}{2}H_{0}\lambda ^{2}+\frac{1}{8}H_{1}\lambda -\frac{1}{256}a_{4}.
\end{eqnarray}
It is easy to derive a second integral $H_{1}$:
\begin{equation*}
H_{1}=p_{2}^{2}q_{1}-p_{1}p_{2}q_{2}-\frac{1}{2}q_{1}^{2}q_{2}^{2}-\frac{1}{8%
}q_{2}^{4}+a_{4}\frac{q_{1}}{4q_{2}^{2}}-\frac{a_{0}}{4}q_{1}q_{2}^{2}+\frac{%
a_{1}}{8}q_{2}^{2}.
\end{equation*}

\section{Exact quasi-periodic solutions}
\label{4}
In this section we give the trajectories of the system under consideration
in terms of Kleinian hyperelliptic functions (see, e.g., \cite%
{ba97,ba07,bel97b,bel97c}), being associated with the real
algebraic curve of genus two (\ref{curvecan1}), which can be also
written in the form
\begin{eqnarray}\label{gen2}
\nu ^{2}=4\prod_{i=0}^{4}(\lambda -\lambda _{i})=4\lambda
^{5}+\sum_{k=0}^{4}\alpha _{k}\lambda ^{k},
\end{eqnarray}
where $\lambda _{i}\neq \lambda _{i}$ are branching points and
\begin{eqnarray}
&&\alpha _{4}=-a_{0},\qquad \alpha _{3}=-\frac{1}{4}a_{1},\qquad \alpha _{2}=%
\frac{1}{2}H_{0},  \notag \\
&&\alpha _{1}=\frac{1}{8}H_{1},\qquad \alpha
_{0}=-\frac{1}{256}a_{4}.
\end{eqnarray}

At all real branching points the closed intervals $[\lambda _{2i-1},\lambda
_{2i}],i=1,2$ will be referred further as \textit{lacunae} \cite{zmnp80,mm75}%
. Let us equip the curve with a homology basis $({\mathfrak{a}}_{1},{%
\mathfrak{a}}_{2};{\mathfrak{b}}_{1},{\mathfrak{b}}_{2})\in H_{1}(K,{\mathbb{%
Z}})$ and fix the basis in the space of holomorphic differentials.

The exact integration of the system (\ref{systemH12}) and
(\ref{systemH22}) reduces to the solution of the Jacobi inversion
problem in the following form
\begin{equation}
\lambda ^{2}-\wp _{22}(\boldsymbol{u})\lambda -\wp _{12}(\boldsymbol{u})=0,
\label{bolza2}
\end{equation}%
that is, the pair $(\mu _{1},\mu _{2})$ is the pair of roots of (\ref{bolza2}%
). So we have
\begin{equation}
\wp _{22}(\boldsymbol{u})=\mu _{1}+\mu _{2},\qquad \wp _{12}(\boldsymbol{u})=-\mu
_{1}\mu _{2}.  \label{b1}
\end{equation}%

Let us introduce finally the \textit{Baker-Akhiezer} function,
which, in the framework of the formalism developed, is expressible
in terms of the Kleinian $\sigma $-function as follows
\cite{bel97b}:
\begin{equation}
\Psi (\lambda ,\boldsymbol{u})=\frac{\sigma \left( \int_{\infty }^{\lambda }{%
\mathrm{d}}\boldsymbol{\ u}-{\mathbf{u}}\right) }{\sigma (\boldsymbol{u})}%
\mathrm{exp}\left\{ \int_{a}^{\lambda }{\mathrm{d}}{\mathbf{r}}^{T}%
\boldsymbol{u}\right\} ,  \label{BAF}
\end{equation}%
where $\lambda $ is arbitrary and $\boldsymbol{u}$ is the Abel image of
arbitrary point $(\nu _{1},\mu _{1})\times (\nu _{2},\mu _{2})\in K\times K$.
It is straightforward to show by the direct calculation,
based on the relations for three and four-index
$\wp$--functions \cite{bel97b}, that $\Psi (\lambda
,\boldsymbol{u})$ satisfies the Schr\"{o}dinger equation
\begin{equation}
\left(\frac{d^{2}}{{d u_{2}}^{2}}-2\wp _{22}(\boldsymbol{u})\right)\Psi
(\lambda ,\boldsymbol{u})=\left( \lambda +\frac{1}{4}\alpha _{4}\right) \Psi
(\lambda ,\boldsymbol{u})  \label{sch}
\end{equation}%
for all $(\nu ,\mu )$. The solutions of (\ref{systemH12}), (\ref{systemH22})
have the following form in terms of Kleinian functions $\wp _{22}(%
\boldsymbol{u}),\wp _{12}(\boldsymbol{u})$ \cite{eekl93,eekt94,k02,beh05}
\begin{equation*}
q_{1}=-2\wp _{22}(\boldsymbol{u}),\qquad q_{2}^{2}=16\wp _{12}(\boldsymbol{u}%
).
\end{equation*}

\section{Elliptic periodic solutions}
\label{5}
In this section we follow the method proposed in \cite{k89,bbeim94,ek94,ee94b},
which allows us to construct periodic solutions of (\ref{systemH12}), (\ref%
{systemH22}) in a straightforward way based on the application of spectral
theory for the Schr\"{o}dinger equation with elliptic potentials \cite%
{amm77,bbeim94,tv90,ve90,ek94,ee94b,sm94,gw95,gw95a,gw96,gw98a,k98,ceek2000,eek00,gvy08}.
We start with the equation (\ref{sch}) for Baker function $\Psi
(\lambda,\boldsymbol{u})$. We assume, without loss of generality,
that the associated curve has the property $\alpha _{4}=0$. To
make this assumption applicable to the initial curve of the system
(\ref{systemH12}), (\ref{systemH22}) being derived from the Lax
representation, we undertake the shift of the spectral parameter
\begin{equation}
\lambda \longrightarrow \lambda -a_{0}/20,  \label{shift1}
\end{equation}

Consider genus $2$ Lam\'e potential ${\mathsf{u}}=6\wp(t+\omega')$
and construct the associated curve
\begin{equation}
\nu^2 = 4 (\lambda^2-3g_2)(\lambda+3e_1) (\lambda+3e_2)(\lambda+3e_3),
\label{curve3}
\end{equation}
The Hermite polynomial ${\mathcal{F}}_3(\wp(t),\lambda)$,
depending on the argument $t+\omega'$, \cite{he12,ww86} associated
to the Lam\'e potential $6\wp(t+\omega')$, which is already
normalized has the form
\begin{equation}
{\mathcal{F}}_3(\wp(t+\omega'),\lambda)=\lambda^{2}-
3\wp(t+\omega')\lambda
 + 9\wp^{2}(t+\omega')-%
\frac{9}{4}g_{2}.  \label{HerPolN}
\end{equation}
Using explicit expression for Hermite polynomial (\ref{HerPolN}) we obtain
the following simple solutions for the system (\ref{HH7}):
\begin{eqnarray}
q_{1}=-6\wp(t+\omega^{\prime }), \qquad
q_{2}^2=-2^4\cdot 3^3\cdot\wp(t+\omega^{\prime})^{2} + 2^2\cdot 3^2 g_{2} ,
\end{eqnarray}
where $a_{0}=0, a_{1}=3\cdot4\cdot 7\cdot g_{2},
a_{4}=-4^{4}\cdot3^{4}\cdot g_{2}g_{3} $. More general solution
with $a_{0}\neq 0$ can be written using the shift (\ref{shift1})
in (\ref{HerPolN}). Then we have
\begin{eqnarray}
&&q_{1}=-6\wp(t+\omega^{\prime })-\frac{1}{5} a_{0}, \\
&&q_{2}^2=-4^2\cdot 3^3\wp(t+\omega^{\prime})^{2}-\frac{12}{5}\wp(t+\omega^{\prime
})a_{0}- \frac{1}{25}a_{0}^2 + 36 g_{2}  ,
\end{eqnarray}
where $a_{1}=3\cdot4\cdot7 \cdot g_{2}-2a_{0}^2/5$ and
\begin{eqnarray}
a_{4}=\frac{1}{3125}a_{0}^5-\frac{84}{125}a_{0}^3 g_{2}-\frac{432}{25}
a_{0}^2 g_{3} +\frac{1728}{5}a_{0} g_{2}^{2}+20736 g_{2} g_{3}.  \notag
\end{eqnarray}
After changing the variables
\begin{eqnarray}
a_{0}=2^2\cdot 5 \cdot\lambda,\qquad \nu=2^8\cdot a_4
\end{eqnarray}
we obtain the curve of genus $2$
\begin{eqnarray}
\nu^2=4 (\lambda^2-3g_2)(\lambda^3-9\lambda g_2 + 27 g_3),
\end{eqnarray}
where $g_{2}, g_{3}$ are elliptic invariants (see for example
\cite{ww86}). The last expression is the another form of
(\ref{curve3}), were $e_j, j=1, 2, 3, \quad e_3\leq e_2\leq e_1$
are real roots of equation $4\lambda^3-g_{2} \lambda -g_{3}$.  The
Weierstrass function $\wp=\wp(t+\omega')$ shifted by half period
$\omega'$ is related to $\mbox{sn}$ Jacobian elliptic function
with modulus $k$ by \cite{Cal}:
\begin{eqnarray}
\wp(t+\omega';g_2,g_3)=\alpha^2 k^2\mbox{sn}^2(\alpha t,k)+ e_{3},
\end{eqnarray}
where $\alpha=\sqrt{e_1 - e_3}$. Using wave height $\alpha$ and
modulus $k=\sqrt{(e_2-e_3)/(e_1-e_3)}$ we have the following
relations (see for example \cite{ww86}):
\begin{eqnarray}
&&e_1=2-k^2,\quad e_2=2k^2-1,\quad
e_3=-(1+k^2),\nonumber \\
&& g_2=2(e_1^2 + e_2^2 + e_3^2)=12(1-k^2+k^4),\nonumber \\
&& g_3=4e_1 e_2 e_3=4(k^2+1)(2-k^2)(1-2k^2).
\end{eqnarray}
From Lax representation for polynomial $F$ we have the following
nonlinear differential equation with spectral parameter $\lambda$
\begin{eqnarray} \label{Feq}
\frac{1}{2} F \ddot{F} -\frac{1}{4} \dot{F}^{2}- (u(t)+\lambda)
F^{2} +\frac{1}{4} \nu^2(\lambda)=0,
\end{eqnarray}
with eigenvalue equations
\begin{eqnarray}
&&\nu^2(\lambda)=4\lambda^{5}-21\lambda^{3} g_{2} +27\lambda
g_{2}^2 + 27 \lambda^2 g_{3}-81 g_{2} g_{3}=0.
\end{eqnarray}

\section{($n+1$)-degrees-of-freedom generalized H\'{e}non-Heiles- type system%
} \label{6}

We consider a generalized H\'{e}non-Heiles-type system with $n+1$ degrees of
freedom \cite{aw92,eekl93,eekt94,t95} :
\begin{eqnarray}
&&\ddot{q}_{0}+3q_{0}^{2}+\frac{1}{2}\sum_{j=1}^{n}q_{j}^{2}+ a_{0}q_{0}%
=0,  \label{systemH1} \\
&&\ddot{q}_{j}+q_{0}q_{j}-\frac{{\mathcal
C}_{j}^{2}}{q_{j}^{3}}-a_{j}q_{j}=0. \label{systemH2}
\end{eqnarray}%
where $j=1,\ldots ,n$. Its Hamiltonian is
\begin{eqnarray}
H =\frac{1}{2}(p_{0}^{2}+\sum_{j=1}^{n}p_{j}^{2})+q_{0}^{3}+\frac{1}{2}%
q_{0}\sum_{j}^{n}q_{j}^{2}+ \frac{1}{2}a_{0} q_{0}^2
-\frac{1}{2}\sum_{j=1}^{n}\left( a_{j}q_{j}^{2} - \frac{{\mathcal C}_{j}^{2}}{%
q_{j}^{2}}\right),  \label{HH}
\end{eqnarray}%
where $q_{0},q_{j},p_{0},p_{j},j=1,\ldots ,n,$ are the canonical
coordinates and momenta, respectively, and $a_{0},{\mathcal
C}_{j}^{2},a_{j},j=1,\ldots ,n,$ are free constant parameters. The
function $H$ for $n=1$ is the Hamiltonian of a classical
integrable H\'{e}non-Heiles system with the additional term
${\mathcal C}_{1}^{2}/q_{2}^{2}$.

Next we will present $(2\times 2)$ matrix Lax representation for the
generalized H\'{e}non-Heiles system (\ref{HH}). The Lax representation has
the form
\begin{equation}
\dot{L}=[M(\lambda ),L(\lambda )],\quad L=\left(
\begin{array}{cc}
V & U \\
W & -V%
\end{array}%
\right) ,\quad M=\left(
\begin{array}{cc}
0 & 1 \\
Q & 0%
\end{array}%
\right)  \label{lax2}
\end{equation}%
where $U,W,Q$ are \cite{eekl93,eekt94,k02}:
\begin{eqnarray}
&&U(t,\lambda )=F(t,\lambda )=a(\lambda )\left( \lambda
+\frac{1}{2}q_{0} + \frac{1}{4} a_{0}
-\frac{1}{16}\sum_{j=1}^{n}\frac{q_{j}^{2}}{\lambda -a_{j}}\right)
,  \notag
\\
&&V=-\frac{1}{2}\dot{F}=a(\lambda )\left( -\frac{1}{4}p_{0}+\frac{1}{16}%
\sum_{j}^{n}\frac{q_{j}p_{j}}{\lambda -a_{j}}\right) ,  \notag \\
&&W=-\frac{1}{2}\ddot{F}+Q F=a(\lambda )\left( \lambda ^{2}-\frac{1}{2}%
q_{0}\lambda + \frac{1}{4}\a_{0} \lambda\right) +  \notag \\
&&+a(\lambda )\left( \frac{1}{4}q_{0}^{2}+\frac{1}{16}%
\sum_{j=1}^{n}q_{j}^{2}\right) +  \notag \\
&&+a(\lambda )\frac{1}{16}\left(
\sum_{j=1}^{n}\frac{p_{j}^{2}}{\lambda
-a_{j}}+\sum_{j=1}^{n}\frac{{\mathcal C}_{j}^{2}}{q_{j}^{2}}\frac{1}{\lambda -a_{j}}%
\right) ,  \notag \\
&&Q(t,\lambda )=\lambda -q_{0}.  \notag
\end{eqnarray}

The corresponding algebraic curve $\det({\mbox
L(t,\lambda)-\frac{\nu}{2} I })=0$ of genus $n+1$ is
\begin{equation}
\nu^{2}=4 a(\lambda)^{2}\left(\lambda^3+ \frac{1}{2} a_{0}
\lambda^2 + \frac{1}{16} a_{0}^2\lambda+\frac{1}{8} H
+\frac{1}{16}\sum_{i=1}^{n} \frac{H_{i}}{\lambda-a_{i}} +
\frac{1}{256} \sum_{i=1}^{n}\frac{{\mathcal
C}_{i}^{2}}{(\lambda-a_{i})^{2}}\right), \label{curveHH}
\end{equation}
where
\begin{eqnarray}
H_{i}=&-&\frac{1}{2}p_{0}q_{i}p_{i} -\left(a_{i}^2 -\frac{1}{2} a_{i} q_{0}+\frac{1}{4}a_{0}a_{i}+\frac{1}{4}q_{0}^2+ \sum_{i=1}^{n}q_{i}^2
\right)q_{i}^{2} +
\frac{1}{2}q_{0}\left(p_{i}^{2}+\frac{{\mathcal C}_{i}^2}{q_{i}^2}\right)
 \notag \\
&+&\frac{1}{4} a_{0} p_{i}^{2}+ a_{i}  p_{i}^{2}+\left(a_{i}+\frac{1}{4} a_{0}\right)\frac{{\mathcal C}_{i}^2}{q_{i}^2} \notag \\
&-&\frac{1}{16}\sum_{k\neq i}\frac{1}{a_{i}-a_{k}}
\left((q_{i}p_{k}-q_{k}p_{i})^{2}- \frac{{\mathcal C}_{i}^2 q_{k}}{q_{i}^2} -%
\frac{{\mathcal C}_{k}^2 q_{i}}{q_{k}^2}\right) ,  \notag \\
\notag \\
&& a(\lambda)=\prod_{i=1}^{n}(\lambda-a_{i}) ,\qquad n\geq 2.
\notag
\end{eqnarray}

Further we will collect the well known facts from finite zone
inverse scattering transform method
\cite{no74,mm75,dmn76,imk77,im75,zmnp80,bbeim94} useful for our
construction of exact solutions of generalized
H\'{e}non-Heiles-type system. Here we follow the Krichever
construction \cite{imk77}. The Baker--Akhiezer (BA)-function
$\psi(t,\lambda)$ of a nonspecial divisor $D$ of degree $n+1$ is
given explicitly by
\begin{eqnarray}
\psi(t,\lambda) = C(P)\,\exp(i\Omega_{1}t)\, \frac{\theta( \mathcal
{A}(P)+\boldsymbol{Z})}{\theta(\boldsymbol{Z})} , \label{BAfun}
\end{eqnarray}
where $\theta(\boldsymbol{v}|B)$ is a Riemann theta function,
$B=(B)_{ij}=\int_{{\mathfrak b}_{i}}\omega_{j}$ is a
Riemann matrix, $\omega_{1}, \omega_{2},\ldots ,\omega_{n+1}$ are
normalized differentials of first kind, $ \mathcal {A}(P)$ is Abel
map $ \mathcal {A}_{k}(P)=\int_{q_{0}}^{P}\omega_{k}$, $q_{0}$ is
arbitrary point on real hyperelliptic Riemann surface
(\ref{curveHH}) denoted by $K$, $\Omega_{1}$ is the normalized
differential of second kind with main parts at $\infty$, $dk$,
$\boldsymbol{Z}=t\boldsymbol{u}+\boldsymbol{Z}_{0}$,
$\boldsymbol{Z}_{0}=- \mathcal {A}(D)-\boldsymbol{K}_{0}$,
$\boldsymbol{K}_{0}$ is the so called Riemann constant vector, the
factor $C(P)$ gives the normalization in (\ref{BAfun}).

The BA-function of nonspecial divisor $D$ is the solution of
\begin{eqnarray}
\left( \frac{d^{2} } {d t^{2}} - u(t)\right)\psi(t,\lambda) =
\lambda \psi(t,\lambda),
\end{eqnarray}
where
\begin{eqnarray}\label{ItsMatveev}
u(t)=2\frac{d^{2}}{d t^{2}}\log{\theta( \boldsymbol {Z})}+
\mbox{const.} \end{eqnarray}

By the Riemann-Roch theorem there exists an nonspecial divisor
$D^{\tau}$ and unique abelian differential $\Omega$ such
that $(\Omega)=D+D^{\tau}-2\infty$ and $\Omega=(1+ O(k^{-2})\,dk$
at $\infty$. There exists a unique function
$\psi^{\tau}(t,\lambda)$ called dual BA--finction of $D^{\tau}$.
The BA-function $\psi^{\tau}(t,\lambda)$ of a nonspecial
divisor $D^{\tau}$ of degree $n+1$ is given explicitly by
\begin{eqnarray}
\psi^{\tau}(t,\lambda) = C^{\tau}(P)\,
\exp(-i\Omega_{1}t )\, \frac{\theta( \mathcal
{A}(P)-\boldsymbol{Z})}{\theta(\boldsymbol{Z})} ,
\end{eqnarray}
The factor $C^{\tau}(P)$ is fixed by the normalization.

Let $u(t)$ be a real finite-gap potential such that the Hill's
operator has only finite number $n$ eigenfunctions defined by
\begin{eqnarray}
&&\psi_{i}=\alpha_{i}\psi(t,p_{i}),\qquad
\tilde{\psi}_{i}=\beta_{i} \psi^{\tau}(t,p_{i}), \nonumber \\
&&\alpha_{i}\beta_{i}= \Res_{p=p_{i}}\Omega\,E,\quad i=1,\ldots ,n
\end{eqnarray}
then the following expansion of $u(t)$ in terms of squared
eigenfunctions have the form
\begin{eqnarray}
u(t) =- \sum_{i=1}^{n}\alpha_{i}\beta_{i}\,
\psi(t,p_{i})\psi^{\tau}(t,p_{i}) + \mbox{const.}
\end{eqnarray}
Let us construct the meromorphic differential
\begin{eqnarray}
\tilde{\Omega} = E\,\psi(t,P)\,\psi^{\tau}(t,P) .
\end{eqnarray}
A straightforward calculations give us
\begin{eqnarray}
&& \Res_{P=p_{0}}\,\tilde{\Omega} = -u(t), \nonumber \\
&& \Res_{P=p_{i}}\,\tilde{\Omega} =-\psi(t,p_{i})
\psi^{\tau}(t,p_{i}) \alpha_{i}\beta_{i} ,
\end{eqnarray}
where $\alpha_{i}\beta_{i}= \Res_{P=p_{i}}\,\tilde{\Omega}\,E$ and
$ \Res_{P=p_{0}}\,\tilde{\Omega}=- \Res_{P=p_{i}}\,\tilde{\Omega}$.

In particular case when $K$ is hyperelliptic Riemann surface
(\ref{curveHH}) or in another form
$\nu^2=4\prod_{i=1}^{2n+3}
(\lambda-\lambda_{i})=R(\lambda ) $. The points of $K$ are pairs $%
P=(\lambda,R)$ and $\lambda(P)$ is the value of the natural projection $%
P\rightarrow\lambda(P)$ of $K$ to the complex projective line $CP^{1}$.

For given nonspecial divisor $D$, there is an unique Baker-Akhiezer (BA)
function $\Psi(t,\lambda)$, such that

(i) the divisor of the poles of $\Psi$ is $D$,

(ii) $\Psi$ is meromorphic on $K\backslash\infty$

(iii) when $P\rightarrow \infty$
\begin{eqnarray}
\Psi(t,P)\exp(-k t) = 1 +\sum_{s=1}^{\infty} m_{s}(t) k^{-s} ,  \label{expan1}
\end{eqnarray}
is holomorphic and $k=\sqrt{\lambda(P)}$ is a local parameter near $P=\infty$%
.

There is a unique function $u(x)$ such that
\begin{equation}
\ddot{\Psi} - u(t)\Psi= \lambda(P)\Psi ,  \label{Hill}
\end{equation}
where $\Psi$ is a BA function. Inserting expansion (\ref{expan1}) into (\ref
{Hill}), we obtain
\begin{eqnarray}
\ddot{\Psi} - 2 \dot{m}(t)\Psi -\lambda(P)\Psi =\exp(kt) O(k^{-1}) ,
\end{eqnarray}
and due to the uniqueness of $\Psi$, we prove (\ref{Hill}), with $u(x)=2
\dot{m}(t)$.

By the Riemann-Roch theorem, there exists a unique differential $\tilde{%
\Omega}$ and a nonspecial divisor $D^{\tau}$ of degree $n$ such
that the
zeros of $\tilde{\Omega}$ are $D + D^{\tau}$ and the expansion at $P=\infty$%
, $\tilde{\Omega}(P)=(1+O(k^{-2})) dk$.

For given nonspecial divisor $D^{\tau}$, there exists a unique dual
Baker-Akhiezer (BA) function such that

(i) the divisor of the poles of $\Psi$ is $D^{\tau}$,

(ii) $\Psi$ is meromorphic on $K\backslash\infty$

(iii) when $P\rightarrow \infty$
\begin{eqnarray}
\Psi^{\tau}(t,P)\exp(-kt) = 1 +\sum_{s=1}^{\infty}\tilde{m}_{s}(t) k^{-s},
\label{expan}
\end{eqnarray}
Fix $\tau$ to be the hyperelliptic involution $P=(\lambda,R)\rightarrow(%
\lambda,-R)$, then we have $D^{\tau}=\tau D$, $\Psi^{\tau}(x,P)=\Psi(x,\tau
P)$. Let $\sum_{i=1}^{n}\mu_{i}(0)$ be the $\lambda$-projection of $D$, and $%
\sum_{i=1}^{n}\mu_{i}(t)$ be the $\lambda$-projection of the zero
divisor of $\Psi(x,P)$. The function $\Psi(t,P)\Psi^{\tau}(t,P)$ is meromorphic on $%
{\bf CP^{1}}$ and the following identity takes place
\begin{equation}
\Psi(x,P)\Psi^{\tau}(t,P)=\frac{F(t,\lambda)}{F(0,\lambda)}  \label{sqfun}
\end{equation}
where $F(t,\lambda)=\prod_{i=1}^{n}(\lambda - \mu_{i}(t))$ and
$\mu_{j}(t)$ satisfies the following system of differential equations
(Kovalevski--Dubrovin equations \cite{dmn76})
\begin{equation}
\frac{d}{dt}\mu_{j}(t) = 2\frac{\sqrt{R(\mu_{j})}} {\prod_{j\neq
k}^{n}(\mu_{j}(t) - \mu_{k}(t))} ,  \label{muSys}
\end{equation}
with initial conditions
\begin{equation}
\mu_{j}(0)\in [\lambda_{2i-1},\lambda_{2i}].
\end{equation}
Equations (\ref{muSys}) are first written by Sonya Kovalevski for genus $n=2$ in relation to
the integrable case of the Kovalevski top and for Korteweg--de Vries hierarchy of equations
by Dubrovin for general $n$ (\cite{dmn76} and references therein).
These equations are useful for numerical calculation
of the polynomial  $F(t,\lambda)$.

Introduce the Wronskian
\begin{eqnarray}
\left\{\Psi(t,P),\Psi^{\tau}(t,P)\right\} &=&\dot{\Psi}(t,P)\Psi^{\tau}(t,P)
-\Psi(t,P)\dot{\Psi}^{\tau}(t,P)  \nonumber\\
& =& \frac{2\sqrt{R(\lambda)}}{\prod_{i=1}^{n}(\lambda -\mu_{i}(0))},
\end{eqnarray}
and the differential $\tilde{\Omega}$ is given explicitly by
\begin{equation}
\tilde{\Omega}(P)=\frac{1}{2}\frac{\prod_{i=1}^{n} (\lambda -
\mu_{i}(0))}{\sqrt{R(\lambda)}}d\lambda .  \label{diff}
\end{equation}
We assume that $E(P)$ is a meromorphic function on $K$ with $n+1$
simple
poles $\infty,p_{1},\ldots,p_{n}$ and at $P\rightarrow\infty$, $%
E(P)=k+\ldots $, and $\tilde{E}(P)$ is meromorphic function with
$n+1$
simple poles $q_{0},q_{1},\ldots ,q_{n}$ and at $P\rightarrow\infty$, $%
\tilde{E}(P)=k^{-1} +\ldots$. We also suppose that the divisors of poles of $%
E(P)$ and $\tilde{E}(P)$ are different from $D$, $D^{\tau}$.
\begin{eqnarray}
\psi(t,\lambda) =\sqrt{F(t,\lambda) }\,
\exp\left( i \int_{0}^{t}\,\frac{\nu(\lambda)}{
F(t^{\prime},\lambda)} d t^{\prime} \right). \label{sBA}
\end{eqnarray}
A brief computation reveals that the BA function solves Hill's
equation
\begin{eqnarray}
\left(\frac{d^2}{d t^2}-u(t)\right)\psi=\lambda \psi,
\end{eqnarray}
and for dual BA function $\psi^{\tau}$ we have
\begin{eqnarray}
\psi^{\tau}(t,\lambda) =&&\sqrt{F(t,\lambda) }\, \exp\left(
-i\int_{0}^{t}\,\frac{\nu(\lambda)}{ F(t^{\prime},\lambda)} d
t^{\prime} \right). \nonumber
\end{eqnarray}
The solutions of the system with Hamiltonian (\ref{HH}) in terms of
Novikov polynomials $F(t,\lambda )$ of degree $n+1$ in spectral
parameter $\lambda$ \cite{no74}, for special points $a_{i}$ $i=1,\ldots ,n$
in closed intervals
$[\lambda_{2i-1},\lambda_{2i}]$, $i=1,\ldots, n$ the functions
are given by
\cite{eekl93,eekt94,k02}
\begin{equation}
q_{0}=-{\mathsf  u}(t),\quad
q_{i}^{2}=16\frac{F(t,a_{i})}{\prod_{k\neq
i}^{n}(a_{i}-a_{k})},\,i=1,\ldots ,n . \label{answern4}
\end{equation}%
${\mathsf  u}(t)$ is the famous Its-Matveev formulae
(\ref{ItsMatveev}), \cite{im75,bbeim94} and the points $a_{i}$ lie
in the lacunae $[\lambda _{2i-1},\lambda _{2i}],i=1,\ldots n,$ for
generalized multidimensional H\'{e}non-Heiles
system and are branch points in the case of the multidimensional H\'{e}%
non-Heiles system (\ref{HH}) with ${\mathcal
C}_{j}^{2}=0,j=1,\ldots ,n$. The solution is real under the choice
of the arbitrary constants $a_{i}, i=1,\ldots ,n$ in such a way,
that the constants $a_{i}, i=1,\ldots, n$ lie in {\it different}
lacunae of Riemann surface $K$.  Then  the constants ${\mathcal
C}_{i}$ are given as
\[{\mathcal C}^2_{i}= \const\cdot\frac{\nu(a_{i})^2}{\left(\prod_ {k\neq i}^{n}
(a_{i}-a_{k})\right)^2}, \label{Cconst}
\]
where $i=1,\ldots ,n$ and $\nu$ is the coordinate of the curve
(\ref{curveHH}).The constants ${\mathcal C}_{i}$ are fixed by the
initial conditions.

Suppose that $F(t,\lambda)={ \mathcal F}(t,\lambda)$, where ${
\mathcal F}$ is Hermite polynomial associated with Lam\'e
potential ${\mathsf u}=(n+1)(n+2)\wp(t+\omega')$. Then the finite
and real solution of the system (\ref{systemH1}), (\ref{systemH2})
is given by (\ref{answern4}) with the Hermite polynomial depending
on the argument $t+\omega'$ (the shift in $\omega'$ provides the
holomorphity of the solution). Consider the potential $12\wp(t)$
and construct the associated curve \cite{he12}
\begin{equation}   \nu^2=4\lambda\prod_{i=1}^{3}
(\lambda^{2}-6e_{i}\lambda+45e_{i}^2-15g_{2}),\label{curve6}
\end{equation}
with branch points given by
\begin{eqnarray}
&&\lambda_{0}=0,\quad \lambda_{1,2}=3\left(1+k^2\pm
2\sqrt{4-7k^2+4k^4}\right),
\nonumber \\
&&\lambda_{3,4}=3\left(1-2k^2\pm 2\sqrt{4-k^2+k^4}\right),
\nonumber \\
&&\lambda_{5,6}=3\left(k^2-2\pm 2\sqrt{1-k^2+4k^4}\right).
\nonumber \\
\end{eqnarray}
or in another form
\begin{eqnarray}
\frac{\nu^2}{4} &=& \lambda^{7}-\frac {63}{2}\lambda^{5} g_2-\frac
{297}{2} \lambda^{4} g_3+\frac {4185}{16}\lambda^{3}
g_2^{2}+\frac {18225}{8}\lambda^{2} g_2 g_3  \nonumber \\ && \qquad + \left( \frac
{91125}{16} g_3^{2}-\frac{3375}{16} g_2^{3} \right) \lambda .
\end{eqnarray}

The Hermite polynomial ${ \mathcal F}(\wp(t),\lambda)$ associated
with the Lam\'e potential $12\wp(t)$ has the form
\begin{eqnarray}
{ \mathcal  F}(\wp(t+\omega'),\lambda)&=&\lambda^{3}-
6\wp(t+\omega')\lambda^2 -3\cdot 5(g_{2}-3\wp(t+\omega')^2)\lambda \nonumber \\
&&\qquad - \frac{3^2\cdot 5^2}{4}(4\wp(t+\omega')^3-g_2\wp(t+\omega')-g_{3}).
\label{HerPol3}
\end{eqnarray}
Then the finite and real solution of the system (\ref{systemH1}),
(\ref{systemH2}) is given by (\ref{answern4}) with the Hermite
polynomial depending on the argument $t+\omega'$. The
constants ${\mathcal C}_{1}$, ${\mathcal C}_{2}$ are given by
\begin{eqnarray}
{\mathcal C}^2_{1}= \const.\cdot \frac{\nu(a_{1})^2}{\left(a_{1}-a_{2}\right)^2},\qquad
{\mathcal C}^2_{2}= \const.\cdot \frac{\nu(a_{2})^2}{\left(a_{2}-a_{1}\right)^2},
\label{Cconst7}
\end{eqnarray}
where $i=1,\ldots ,n$ and $\nu$ is the coordinate of the curve
(\ref{curve6}).

Next we list the periodic solutions of the system
(\ref{systemH1}), (\ref{systemH2}) for $n=2$ and for genus $3$,
where ${\mathcal C}_{1}=0$ and ${\mathcal C}_{2}=0$. The $(2n+3)$
Lam\'e polynomials of order $n+1$ are solutions of
\begin{eqnarray}
\frac{d^2 E_{i}}{d t^2}
+\left((n+1)(n+2)k^2\mbox{dn}^2(\alpha t) - \lambda_{i}\right) E_{i}=0.
\end{eqnarray}
For $n=2$ we introduce the following eigenfunctions
$\mbox{E}^{(3)}_{i},\,i=1,\ldots 7$ and eigenvalues $\lambda_{i}$
given in table \ref{tab:2}, $\lambda_1<\lambda_2<\ldots
<\lambda_7$ and the results are collected in Table \ref{tab:2}, see for example \cite{psv99}.
\begin{table}[h]
\caption{\label{tab:2}}
\begin{tabular}{|l|l|l|} \hline
1  & $\mbox{E}_{1}^{(3)}=\mbox{sn}(\alpha t,k)\left(
\mbox{dn}^2(\alpha t,k)+C_{1}^{(3)}\right)$ &
$\lambda_{1}^{(3)}=7-5 k^2-2\sqrt{4-7k^2+4k^4}$ \\
\hline 2 & $\mbox{E}_{2}^{(3)}=\mbox{cn}(\alpha t,k)\left(
\mbox{dn}^2(\alpha t,k)+C_{2}^{(3)}\right)$ & $
\lambda_{2}^{(3)}=7-2 k^2-2\sqrt{4-k^2+k^4} $  \\
\hline 3 & $\mbox{E}_{3}^{(3)}=\mbox{dn}(\alpha t,k)\left(
\mbox{dn}^2(\alpha t,k)+C_{3}^{(3)}\right)$ &
$ \lambda_{3}^{(3)}=5(2-k^2)-2\sqrt{1-k^2+4k^4}$   \\
\hline 4 & $\mbox{E}_{4}^{(3)}=\mbox{sn}(\alpha
t,k)\mbox{cn}(\alpha t,k)\mbox{dn}(\alpha t,k)$ &
$ \lambda_{4}^{(3)}=4(2-k^2) $   \\
\hline 5 & $\mbox{E}_{5}^{(3)}=\mbox{sn}(\alpha t,k)\left(
\mbox{dn}^2(\alpha t,k)+C_{5}^{(3)}\right)$ & $
\lambda_{5}^{(3)}=7-5 k^2+2\sqrt{4-7k^2+4k^4}$   \\
\hline 6 & $\mbox{E}_{6}^{(3)}=\mbox{cn}(\alpha t,k)\left(
\mbox{dn}^2(\alpha t,k)+C_{6}^{(3)}\right)$ & $
\lambda_{6}^{(3)}=7-2 k^2+2\sqrt{4-k^2+k^4} $  \\
\hline 7 & $\mbox{E}_{7}^{(3)}=\mbox{dn}(\alpha t,k)\left(
\mbox{dn}^2(\alpha t,k)+C_{7}^{(3)}\right)$ &
$ \lambda_{7}^{(3)}=5(2-k^2)+2\sqrt{1-k^2+4k^4}$    \\
\hline
\end{tabular}
\end{table}
\begin{center}
\begin{table}[h]
\caption{\label{tab:3}}
\begin{tabular}{|l|l|} \hline
1 &
$C_{1}^{(3)}=\frac{1}{5}(2k^2-3-\sqrt{4-7k^2+4k^4})$ \\
\hline 2& $
C_{2}^{(3)}=\frac{1}{5}(k^2-3-\sqrt{4-k^2+k^4}) $  \\
\hline 3 &
$ C_{3}^{(3)}=\frac{1}{5}(2k^2-4-\sqrt{1-k^2+4k^4})$   \\
\hline 4 &
   \\
\hline 5 &  $
C_{5}^{(3)}=\frac{1}{5}(2k^2-3+\sqrt{4-7k^2+4k^4})$   \\
\hline 6 & $
C_{6}^{(3)}=\frac{1}{5}(k^2-3+\sqrt{4-k^2+k^4}) $  \\
\hline 7 &
$ C_{7}^{(3)}=\frac{1}{5}(2k^2-4+\sqrt{1-k^2+4k^4})$    \\
\hline
\end{tabular}
\end{table}
\end{center}
For convenience we present solutions in the following form
\begin{eqnarray}
q_{0}=12 \tilde{C}_{0}k^2\mbox{dn}^2(\alpha t),\qquad
q_{1}^{(i)}=\tilde{C}^{(i)}_{1}\mbox{E}_{i}^{(3)},\qquad
q_{2}^{(j)}=\tilde{C}_{2}^{(j)}\mbox{E}^{(3)}_{j},\qquad i\neq
j=1,\ldots 7
\end{eqnarray}
where the constants $\tilde{C}_{0}$,
$\tilde{C}^{(i)}_{1},\tilde{C}^{(i)}_{2}$ are fixed by initial
conditions. The same procedure is possible for general $n$.

The Hermite polynomial ${ \mathcal  F}(\wp(t),\lambda)$ associated
with the Lam\'e potentials  can be written in a different form,
useful for applications:
\begin{eqnarray}
{ \mathcal  F}(\wp(t),\lambda)=\sum_{k=0}^{n}
A_{k}(\lambda)\wp(t)^{n-k}.
\end{eqnarray}
For example for the genus $4$ Lam\'e potential $20\wp(t)$ we have
\begin{eqnarray}
&&A_{0}=11025,\qquad A_{1}=-1575\lambda,\qquad
A_{2}=135\lambda^2-\frac{6615}{2}g_2 \nonumber \\ &&
A_{3}=-10\lambda^3+\frac{1855}{4}\lambda g_{2} -2450 g_{3}
\nonumber \\&& A_{4}=\lambda^4-\frac{113}{2}\lambda^2 g_{2}
+\frac{3969}{16} g_{2}^2 +\frac{1925}{4}\lambda g_{3}. \nonumber
\end{eqnarray}
or in explicit form we have the Hermite polynomial ${ \mathcal
F}(\wp(t+\omega'),\lambda)$ associated to the Lam\'e potential
$20\wp(t+\omega')$ can be written as
\begin{eqnarray}
{ \mathcal
F}(\wp(t+\omega'),\lambda)&=&11025\wp(t+\omega')^4-1575\wp(t+\omega')^3
\lambda  \nonumber\\ &&
+ (135\lambda^2-\frac{6615}{2}g_{2})\wp(t+\omega')^2 \nonumber
\\&& +(-10\lambda^3+ \frac{1855}{4}\lambda g_{2}-2450 g_{3})\wp(t+\omega')
\nonumber \\&& + \lambda^4-\frac{113}{2}\lambda^2 g_{2}+
\frac{3969}{16}g_{2}^2+\frac{195}{4}\lambda g_{3}. \label{HerPol4}
\end{eqnarray}

For  $n=3$ and genus four, the Lam\'e curve  have the following form
\begin{eqnarray}
\nu^2=4\left(\prod_{l=1}^{3} (\lambda^{2}+10 e_{l}\lambda - 35
e_{l}^2-7 g_{2})\right)(\lambda^3-52\lambda g_{2} + 560 g_{3}).
\nonumber
\end{eqnarray}
or in another form convenient for practical use we have
\begin{eqnarray}
\frac{\nu^2}{4}&=&\lambda^9-\frac{231}{2}\lambda^7
g_{2}+\frac{2145}{2}g_{3}\lambda^6+ \frac{63129}{16}\lambda^5
g_{2}^2-\frac{518505}{8}g_{2} g_{3}\lambda^4 \nonumber
\\&&+\left(-\frac{563227}{16}g_{2}^3+\frac{4549125}{16}g_{3}^2\right)
\lambda^3+\frac{991515}{2} g_{3}g_{2}^2\lambda^2 \nonumber\\ && +
\left(\frac{361179}{4}g_{2}^4-\frac{5273625}{4}g_{2}g_{3}^2\right)\lambda
 -972405g_{3}g_{2}^3-1500625g_{3}^3.\label{lame4}
\end{eqnarray}

\section{Summary and conclusions} \label{7}

We approached the most general
class among the three classes of integrable Henon- Heiles-type
systems identified in \cite{BSV}.

We provided exact quasi-periodic
solutions for this class of systems, expressing these orbits via
Kleinian hyper-elliptic functions.

Applying the spectral theory for
the Schrodinger equation to elliptic potentials, we pointed out
elliptic periodic solutions.

To obtain these results, we resorted
to various mathematical methods, characteristic to both physics
and astronomy.

We applied these methods and results to a
generalized Henon-Heiles-type system with $(n+1)$ degrees of
freedom, pointing out the exact solutions in this case.

By emphasizing the exact solutions for the integrable class of
systems under consideration, our paper contributes to a better
understanding of the generalized Henon- Heiles-type problem.

\end{document}